\documentclass[a4paper,11pt]{article}
\pdfoutput=1 

\usepackage{jheppub,bm,color} 
                     
\usepackage[T1]{fontenc} 

\newcommand{\be}{\begin{equation}}      
\newcommand{\ee}{\end{equation}}      
\newcommand{\bea}{\begin{eqnarray}}      
\newcommand{\eea}{\end{eqnarray}}    
     
\newcommand{\rt}[1]{{}}

\newcommand{\Tr}{\,\textrm{Tr}\,}

\newcommand{\SU}{\,\textrm{SU}\,}
\newcommand{\U}{\,\textrm{U}\,}

\newcommand{\const}{\,\textrm{const.}\,}

\title{Functional renormalization group approach to color superconducting phase transition} 

\author{Gergely Fej\H{o}s$^{1,2,3}$ and Naoki Yamamoto$^{1,2}$}

\affiliation{$^1$Department of Physics, Keio University, Yokohama 223-8522, Japan}
\affiliation{$^2$Research and Education Center for Natural Sciences, Keio University, Yokohama 223-8521, Japan}
\affiliation{$^3$Institute of Physics, E\"otv\"os University, 1117 Budapest, Hungary}

\emailAdd{fejos@keio.jp}
\emailAdd{nyama@rk.phys.keio.ac.jp}

\abstract{We investigate the order of the color superconducting phase transition using the functional renormalization group approach. We analyze the Ginzburg-Landau effective theory of color superconductivity and more generic scalar $\SU(N_{\rm c})$ gauge theories by calculating the $\beta$ function of the gauge coupling in arbitrary dimension $d$ based on two different regularization schemes. We find that in $d=3$, due to gluon fluctuation effects, the $\beta$ function never admits an infrared fixed point solution. This indicates that, unlike the ordinary superconducting transition, color superconductivity 
can only show a first-order phase transition.}

\begin{document} 
\maketitle
\flushbottom

\section{Introduction}
Color superconductivity is a phenomenon that is displayed by degenerate quark matter at high densities, well above nuclear saturation density. 
In contrast to the conventional metallic superconductivity caused by phonon-mediated attractive interactions, where the fundamental Coulomb interaction between electrons itself is repulsive, the gluon exchange interaction between quarks can be attractive and directly leads to the instability of the Fermi surface against the formation of Cooper pairs of quarks. Owing to the color and flavor degrees of freedom in QCD, a variety of color superconducting phases are expected to appear depending on the temperature $T$ and baryon chemical potential $\mu$, such as the color-flavor locked (CFL) phase \cite{alford99} and two-flavor color superconductivity (2SC) phase \cite{bailin83} among others; see, e.g. \cite{alford08} for a review. 

Previously, it was argued that the order of the color superconducting phase transition is of first order. It was shown that even though the Ginzburg-Landau free energy of color superconductivity \cite{iida02,iida02b} predicts the transition to be of second order, once thermal gluon fluctuations are taken into account, the Coleman-Weinberg mechanism most likely drives the system away from critical behavior \cite{pisarski99,matsuura04,giannakis04,noronha06} (though a second-order line in a smaller $\mu$ region cannot be entirely ruled out \cite{pisarski00}). Similar arguments were also applied to ordinary superconductivity \cite{halperin74}, where it is well known that they can completely fail. It is believed that vortex fluctuations are of huge importance around the critical temperature in the type-II regime \cite{kleinert82,kleinert06}, and perturbative calculations of the free energy, such as the one-loop result of the Coleman-Weinberg potential, become invalid. Likewise, the $\epsilon$ expansion of the $\beta$ functions can also be applied to ordinary superconductors for $\epsilon=1$, where it is also known to fail \cite{folk96}. It was only shown recently in an analytic fashion via the functional renormalization group (FRG) framework that, ordinary superconductivity does possess nontrivial charged (i.e. nonzero gauge coupling) fixed points \cite{fejos16,fejos17} describing the possibility of a second-order transition in the system, in agreement with Kleinert's duality argument \cite{kleinert82} and Monte-Carlo simulations \cite{mo02}; for reviews of FRG, see, e.g. \cite{berges00,kopietz,delamotte12}.

The main motivation of this study is to perform a more thorough investigation of the phase transition order of color superconductivity in terms of the FRG approach. We wish to perform a similar analysis already done in the Abelian-Higgs model \cite{fejos16,fejos17}, here generalizing it to non-Abelian gauge theories. It is important to stress that, through the FRG method, one is able to access all $\beta$ functions directly in three dimensions ($d=3$), without using the $\epsilon$ expansion.%
\footnote{The phase transition of a non-Abelian gauge theory with fundamental bosons was previously studied using the $\epsilon$ expansion in \cite{das18} in the context of condensed matter physics.} 
Our focus will be on the Ginzburg-Landau effective theory of color superconductivity and more generic scalar $\SU(N_{\rm c})$ gauge theories in $d$ Euclidean space dimensions.

In this paper we will calculate the RG flow of the gauge coupling, and argue that for $N_{\rm c}=3$ and $N_{\rm f} \leq 55$ (where $N_{\rm c}$ and $N_{\rm f}$ are the numbers of colors and flavors, respectively), no IR stable fixed points can exist unlike for ordinary superconductivity, irrespectively of the concrete form of the scalar potential, due to strong gluon fluctuations. This conclusion can be reached even in the nonperturbative strongly-coupled regime under the leading-order approximation%
\footnote{This is also called the local potential approximation' (LPA') in the literature.}
of the derivative expansion of the quantum effective action (with nontrivial wavefunction renormalizations). This approximation, in principle, limits the method to systems with the small anomalous dimensions \cite{berges00}.

Applying the FRG to gauge theories is not straightforward. Being a generalization of the Wilsonian idea of the RG, it includes by construction a momentum cutoff scale $k$. This explicitly breaks local gauge symmetry, and one is faced with the problem of validity of such an approach. The problem has been tackled by several methods. The background field approach, in which gauge symmetry still exists in terms of background field transformations, has been a popular scheme \cite{reuter94,reuter94b,bergerhoff96,bergerhoff96b,gies02}. Manifestly gauge invariant constructions of the renormalization group flows have also been invented without referring to the Fadeev-Popov method \cite{morris01,arnone02,arnone03,arnone07}. By introducing macroscopic gauge fields, another version of a gauge-invariant flow equation has appeared recently \cite{wetterich18}. A new method argues that gauge (or BRST) symmetry is not necessarily broken by the presence of a cutoff, once one is able to make the cutoff function part of the gauge-fixing condition \cite{asnafi18}. In \cite{igarashi19} one finds a way to accommodate the quantum master equation for BRST symmetry with the RG flow equation. 

In spite of having theoretical successes of implementing gauge invariance into the RG concept, they are rather complicated and not easily accessible for practical computations. The usual functional integral approach with covariant gauge fixing is still the simplest way to go. In this paper we also stick to this method, and tackle gauge symmetry violation at the level of the approximation. We will see that the incompatibility of gauge symmetry and the presence of a momentum cutoff can be (partially) cured via appropriate choice(s) of gauge-fixing parameters, which need to be adjusted in accordance with the symmetry group and the matter content.

The paper is organized as follows. In section \ref{sec:GL}, we briefly review the Ginzburg-Landau theory of color superconductivity.
In section \ref{sec:basics}, we recall the basics of the FRG method and the employed approximations. Section \ref{sec:diagram} is the main part, where we perform all calculations using the diagrammatic technique that leads to the RG flow of the gauge coupling. Section \ref{sec:vertex} is devoted for analyzing how the results are affected once one implements on top of momentum space cutoff in the loop integrals, vertex regularizations. The reader finds the summary in section \ref{sec:conclusion}.

\section{Ginzburg-Landau theory of color superconductivity}
\label{sec:GL}
We first briefly review the Ginzburg-Landau (GL) theory of color superconductivity \cite{iida02,iida02b}. Generically, the idea of the GL theory is to expand the thermodynamic potential of a system in terms of an order parameter and its derivatives near the second-order or weak first-order phase transition. In the case of color superconductivity in QCD with $N_{\rm c} = 3$ and $N_{\rm f} = 3$, the order parameter is the scalar field $\phi^{\gamma}_{n}$ defined by
\bea
\langle \psi_{l}^{\alpha} C \gamma_5 \psi_m^{\beta} \rangle \sim \epsilon^{\alpha \beta \gamma} \epsilon_{l m n} \phi^{\gamma}_{n}\,,
\eea
where $\psi$ is the quark field with $\alpha, \beta, \gamma$ and $l, m, n$ being color and flavor indices, respectively, and $C$ is the charge conjugation operator. Here, the formation of the quark-quark pairing is assumed to be in the parity-even, s-wave, and color-flavor antisymmetric channel. The assumption that the pairing takes place in the color antisymmetric channel is indeed justified at sufficiently high density, where the dominant one-gluon exchange interaction at weak coupling is attractive. The instanton-induced interaction, which may be relevant at lower density region, is also attractive in this channel, and so this assumption is reasonable independently of the density. Then, for the s-wave pairing, the flavor antisymmetry follows from the Fermi statistics of quarks.

Let us consider massless QCD. The thermodynamic potential $V(\phi)$ can be expanded in terms of $\phi$, such that each term respects the symmetries of massless QCD, ${\cal G} \equiv \SU(3)_{\rm c} \times \SU(3)_{\rm L} \times \SU(3)_{\rm R} \times \U(1)_{\rm B}$, as
\bea
\label{GL}
V(\phi) = V_0 + \alpha \Tr(\phi^{\dag} \phi) + \beta_1 \big[\Tr(\phi^{\dag} \phi) \big]^2 + \beta_2 \Tr \big[(\phi^{\dag} \phi)^2 \big] + \cdots\,,
\eea
where the coefficients $\alpha$, $\beta_1$, $\beta_2$ are some $T$ and $\mu$-dependent parameters that cannot be fixed by the symmetries alone. At asymptotically high density, these parameters can be computed by the weak-coupling computation \cite{iida02}, but their detailed expressions will be irrelevant to our following discussions. The two typical ground states of high-density quark matter are the so-called color-flavor locked (CFL) phase \cite{alford99} and two-flavor color superconductivity (2SC) phase \cite{bailin83} of the forms
\bea
\langle \phi^{\gamma}_n \rangle = \left \{
\begin{array}{l}
\Delta \delta^{\gamma}_n \quad \ \ ({\rm CFL}) \\
\Delta \delta^{\gamma}_3 \delta^3_n \quad  ({\rm 2SC})
\end{array}
\right.\,,
\eea
respectively, where $\Delta$ is the gap parameter characterizing the magnitude of the pairing. Whether the ground state becomes the CFL, 2SC, or possible other ground states depends on the form of $V(\phi)$ (i.e. the parameters $\alpha$, $\beta_1$, $\beta_2$).

In the following, we will consider generic scalar $\SU(N_{\rm c})$ gauge theories (which includes the GL theory of color superconductivity with $N_{\rm c}=3$ above) in $d$-dimensional Euclidean space:%
\footnote{In the case of the GL theory of color superconductivity, the kinetic term for $\phi$ takes the form, $\kappa \Tr \big[ (D_i \phi)^{\dag} (D_i \phi) \big]$, with some parameter $\kappa$. Nonetheless, we can choose the normalization $\kappa = 1$ by the rescaling $\phi \rightarrow \kappa^{-1/2} \phi$. (Accordingly, the coefficients $\alpha$, $\beta_{1,2}$ in the potential $V(\phi)$ also change, but the details of $V(\phi)$ will in any case be irrelevant below.)}
\bea
{\cal L}_0 =- \frac{1}{2}\Tr (F_{ij}F_{ij})+ \Tr \big[ (D_i \phi)^{\dag} (D_i \phi) \big] + V(\phi).
\eea
Here, $\phi \equiv \phi_{\gamma}^n$ is the scalar field that have generic number of flavors ($n = 1, 2, \dots, N_{\rm f}$) and colors ($\gamma = 1, 2, \dots, N_{\rm c}$), $D_i = \partial_i + {\rm i}g A_i$ is the covariant derivative with $A_i = A_i ^a \hat T^a$ being the $\SU(N_{\rm c})$ gauge field and $\hat{T}^a$ the generators of $\SU(N_{\rm c})$ in the antifundamental representation,  $F_{ij}\equiv F_{ij}^a\hat{T}^a$ is the field strength of the gauge field, $F_{ij}^a=\partial_i A_j^a - \partial_j A_i^a + g f^{abc}A_i^bA_j^c$, and $V(\phi)$ is a generic potential term. In the standard covariant gauge, the Lagrangian is reformulated as
\bea
\label{Eq:Lag}
{\cal L}&=&\frac12A_i^a\delta^{ab}\Big(-\partial^2\delta_{ij}+(1-\xi^{-1})\partial_i \partial_j\Big) A_j^b+\bar{c}^a(-\partial^2\delta^{ac}-gf^{abc}\partial_i A_i^b)c^c - \phi^{\dagger n}_\gamma \partial^2 \phi^n_\gamma + V(\phi) \nonumber\\
& &+ {\rm i}gA_i^a\Big(\partial_i\phi_\gamma^\dagger (\hat{T}^a\phi_\gamma)-(\hat{T}^a\phi_\gamma)^\dagger \partial_i \phi_\gamma \Big) +g^2f^{abe}f^{cde} A_i^a A_i^b (\hat{T}^c\phi_\gamma)^\dagger(\hat{T}^d\phi_\gamma) \nonumber\\
& &+gf^{abc}\partial_i A_j^a A_i^b A_j^c+\frac{g^2}{4}f^{abe}f^{cde}A_i^aA_j^bA_i^cA_j^d\,,
\eea
where $\bar{c}^a$ and $c^a$ ($a=1, 2, \dots, N_{\rm c}^2-1$) are the ghosts and $f^{abc}$ are the totally antisymmetric structure constants of $\SU(N_{\rm c})$. Note that, in (\ref{Eq:Lag}) electromagnetism is not included for simplicity, even though in principle the photon mixes with one of the gluons in the color superconducting phases. The reason is that it turns out that at the level of the forthcoming analysis, the inclusion of the $\U(1)$ electromagnetic gauge field cannot change the results neither qualitatively nor quantitatively.

The question we would like to ask is the order of the phase transition between the ordered phase ($\phi \neq 0$) and disordered phase ($\phi = 0$) in the aforementioned class of theories.

\section{Basics of the functional renormalization group}
\label{sec:basics}
The framework employed in this study is the functional renormalization group (FRG). In the core of the formalism lies the scale-dependent quantum effective action, $\Gamma_k$, which incorporates all the fluctuations of a field theory beyond a momentum scale $k$. That is, $\Gamma_k$ is infrared (IR) regulated via a regulator function ${\cal R}_k$ (which, in principle, is a matrix in the internal space of fields), added to the classical Lagrangian as a momentum-dependent mass term. Given that the latter is chosen appropriately, it provides large mass to low momentum modes, making them impossible to propagate, thus, in this sense, it generalizes the idea of the Wilsonian renormalization group to the level of the effective action. There are numerous ways to fix the ${\cal R}_k$ matrix, and we will come back to this point during the explicit calculations below.

The $\Gamma_k$ function obeys the so-called flow equation \cite{wetterich92},
\bea
\label{Eq:flow}
k\partial_k \Gamma_k = \frac12 \int k\tilde{\partial}_k \Tr \log \Big(\Gamma_k^{(2)}+{\cal R}_k\Big)\,,
\eea
where $\Gamma_k^{(2)}$ is the second functional derivative matrix of $\Gamma_k$ with respect to all the fields, and the differential operator $k\tilde{\partial}_k$ acts only on ${\cal R}_k$ (but not on $\Gamma_k^{(2)}$). The log function has to be considered in the functional sense (i.e. via its series representation), and each term in the integrals can be considered either in coordinate or momentum space. The trace operation has to be taken over the internal space of fields. The $\beta$ function of any coupling is defined similarly as in the Wilsonian renormalization group: one extracts in the $\Gamma_k$ effective action the momentum-independent part of each proper vertex, and keeps track of their scale evolution.

We are interested in the flow of the gauge coupling in the family of theories defined in (\ref{Eq:Lag}), especially for $N_{\rm c}=N_{\rm f}=3$. After the usual rescaling of the dynamical variables and coupling(s), 
\bea
\label{Eq:rescale}
A_i &\rightarrow& Z_A^{1/2} A_i, \qquad \phi \rightarrow Z_\phi^{1/2}\phi, \qquad c \rightarrow Z_c^{1/2}c, \qquad g \rightarrow Z_g g\,,
\eea
with $Z_i\equiv 1+\delta Z_i$ ($i=A,\phi,c,g$), and redefinitions via counterterms, 
\bea
Z_g Z_c Z_A^{1/2} &\Rightarrow& 1+\delta_{A\bar{c}c}, \quad Z_gZ_{\phi}Z_A^{1/2} \Rightarrow 1+\delta_{\phi^2A}, \qquad Z_g^2Z_\phi Z_A \Rightarrow 1+\delta_{\phi^2 A^2},  \nonumber\\
Z_gZ_A^{3/2} &\Rightarrow& 1+\delta_{3g}, \qquad Z_g^2Z_A^2 \Rightarrow 1+\delta_{4g}\,,
\eea
there are various ways to define the $\beta$ function of the gauge coupling. The Slavnov-Taylor identities guarantee that the following expressions are equal and they serve as identical definitions (at leading order):
\begin{itemize}
\item $-g \mu\partial_\mu (\delta_{\phi^2 A}-\delta Z_\phi - \delta Z_A/2)$
\item $-g \mu\partial_\mu (\delta_{\phi^2 A^2}-\delta Z_{\phi}-\delta Z_A)/2$
\item $-g \mu \partial_\mu (\delta_{3g}-3\delta Z_A/2)$
\item $-g \mu \partial_\mu (\delta_{4g}/2-\delta Z_A)$
\item $-g \mu \partial_\mu (\delta_{A\bar{c} c}-\delta Z_c -\delta Z_{A}/2)$
\end{itemize}
Here $\mu$ is the renormalization scale and we think of the counterterms as quantities evaluated in dimensional regularization close to $d=4$.

In the presence of a momentum space regulator, however, gauge invariance and the Slavnov-Taylor identities are explicitly broken, and thus, the FRG versions of the listed definitions may not be equal. We do not intend to calculate all of them for comparison, but choose the diagrammatically simplest, the last one. This choice is also motivated from the point of view of the non-Abelian version of the QED Ward identity stating that $\delta_{\phi^2A}-\delta Z_\phi$ (and similarly $\delta_{\phi^2A^2}-\delta Z_\phi$) is independent of the matter sector \cite{peskin}. One expects that due to the momentum space regulator, cancellations of matter contributions might not survive, but using the last definition, no scalars appear in the corresponding diagrams of $\delta_{A\bar{c}c}-\delta Z_c$, and one needs not worry about such anomalous terms. As shown in \cite{fejos19}, this is of particular importance when searching for IR fixed points.

Under the leading-order approximation of the derivative expansion, the scale-dependent quantum effective action based on (\ref{Eq:Lag}) reads
\bea
\label{Eq:GammakAns}
\Gamma_k &=& \int_x \bigg[\frac{Z_{A,k}}{2}A_i^a\delta^{ab}\Big(-\partial^2\delta_{ij}+(1-\xi_k^{-1})\partial_i \partial_j\Big)A_j^b+Z_{c,k}\bar{c}^a(-\partial^2\delta^{ac}-Z_{g,k}Z_{A,k}^{1/2}gf^{abc}\partial_i A_i^b)c^c \nonumber\\
&&\hspace{0.7cm}-Z_{\phi,k}\phi^{\dagger n}_\gamma\partial^2 \phi^n_\gamma + V_k(\phi) + {\rm i} Z_{g,k}Z_{A,k}^{1/2}Z_{\phi,k}gA_i^a \Big(\partial_i\phi_\gamma^\dagger (\hat{T}^a\phi_\gamma)-(\hat{T}^a\phi_\gamma)^\dagger \partial_i \phi_\gamma \Big) \nonumber\\
&&\hspace{0.7cm}+Z_{g,k}^2Z_{A,k}Z_{\phi,k}g^2f^{abe}f^{cde} A_i^a A_i^b (\hat{T}^c\phi_\gamma)^\dagger(\hat{T}^d\phi_\gamma) +Z_{g,k}Z_{A,k}^{3/2}gf^{abc}\partial_i A_j^a A_i^b A_j^c \nonumber\\
&&\hspace{0.7cm}+Z_{g,k}^2Z_{A,k}^2\frac{g^2}{4}f^{abe}f^{cde}A_i^aA_j^bA_i^cA_j^d\bigg]\,.
\eea
Before moving on to the calculations, we specify the regulator. For each mode of $\Phi\equiv (A_i^a, c^a, \phi^n_\gamma)$, we add the optimal regulator\footnote{The optimal regulator guarantees that e.g. the convergence radius in an amplitude expansion of the effective potential is maximal.} $R_k(q)=(k^2-q^2)\Theta(k^2-q^2)$ \cite{litim:2001} as
\bea
\int \Phi^\dagger {\cal R}_k \Phi &=& \frac{Z_{A,k}}{2} \int_q A_i^a(q)A_j^a(-q) \Big(\delta_{ij}+\hat{q}_i\hat{q}_j(1-\xi_k^{-1})\Big) R_k(q) \nonumber\\
& &+Z_{c,k}\int_q \bar{c}^a(q)c^a(-q) R_k(q)+Z_{\phi,k} \int_q \phi^{\dagger n}_\gamma(q)\phi^n_\gamma(-q)R_k(q)\,.
\eea
Since the right-hand side of (\ref{Eq:flow}) is a one-loop expression, one needs to find all the one-loop diagrams built up from vertices and regulated propagators of (\ref{Eq:GammakAns}). As announced already, the $\beta$ function of the gauge coupling will be determined through the gauge-antighost-ghost vertex. Based on (\ref{Eq:rescale}), introducing the flowing gauge coupling as $g_k=gZ_{g,k}$ (with $g$ being the bare coupling) and its dimensionless version as $\bar{g}_k=k^{d-4}g_k$, we have
\bea
\label{Eq:betagdef}
\beta(g)&\equiv &k\partial_k \bar{g}_k = (d-4)\bar{g}_k+gk^{d-4}k\partial_k Z_{g,k}=(d-4)\bar{g}_k+gk^{d-4}k\partial_k \Bigg(\frac{Z_{g,k}Z_{A,k}^{1/2}Z_{c,k}}{Z_{A,k}^{1/2}Z_{c,k}}\Bigg)\nonumber\\
&=&(d-4)\bar{g}_k+\frac{\bar{g}_k}{Z_{g,k}Z_{A,k}^{1/2}Z_{c,k}}k\partial_k (Z_{g,k}Z_{A,k}^{1/2}Z_{c,k})-\frac{\bar{g}_k}{Z_{c,k}}k\partial_k Z_{c,k}-\frac{\bar{g}_k}{2Z_{A,k}}k\partial_k Z_{A,k}\,. \nonumber\\
\eea
In what follows, we calculate $k\partial_k (Z_{g,k}Z_{A,k}^{1/2}Z_{c,k})$, $k\partial_k Z_{c,k}$, and $k\partial_k Z_{A,k}$, respectively, to obtain $\beta(g)$.

\section{Diagrammatics}
\label{sec:diagram}
The Feynman rules can be read off from (\ref{Eq:GammakAns}). The wiggly, straight, and dotted lines correspond to gauge, scalar and ghost propagators, respectively. First, we calculate the gauge-antighost-ghost vertex, then the gauge and ghost wavefunction renormalizations. The evaluation of common integrals in the following calculations can be found in the appendix.

\subsection{Gauge-antighost-ghost vertex}
Diagrammatics tells us that
\bea
\label{Eq:vertex}
p_igf^{abc}k\partial_k (Z_{g,k}Z_{A,k}^{1/2}Z_{c,k})=k\tilde{\partial}_k\Bigg( \hspace{-6.0cm}\includegraphics[bb = -140 528 250 0,scale=0.5,angle=0]{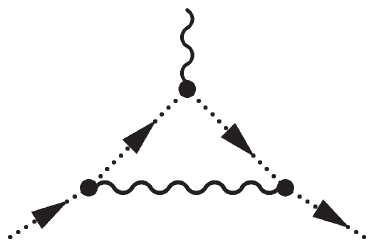}\hspace{1cm}+\hspace{-6.0cm}\includegraphics[bb = -140 528 250 0,scale=0.5,angle=0]{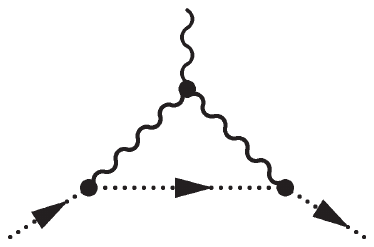}\hspace{1.1cm}\Bigg)\,.
\eea
The two diagrams need to be expanded up to ${\cal O}(p)$ as
\vspace{0.2cm}
\bea
\label{Eq:vertexd1}
k\tilde{\partial}_k\Bigg( \hspace{-6.0cm}\includegraphics[bb = -140 528 250 0,scale=0.5,angle=0]{vertex1.pdf}\hspace{1cm}\Bigg)&=&g^3Z_{g,k}^3Z_{A,k}^{1/2}Z_{c,k}\int_q k\tilde{\partial}_k \frac{1}{q_R^2}\Big(\delta_{mn}-(1-\xi_k)\frac{q_mq_n}{q^2}\Big)\frac{1}{((q-p)_R^2)^2}\nonumber\\
&&\hspace{2.5cm}\times (p-q)_m (q-p)_i p_n f^{eoc}f^{fbe}f^{aof}\nonumber\\
&=&p_if^{abc}g^3Z_{g,k}^3Z_{A,k}^{1/2}Z_{c,k}C_2(A)\Omega_d k^{d-4}\frac{-3\xi_k}{d(d+2)}+{\cal O}(p^2)\,,
\eea
\bea
\label{Eq:vertexd2}
k\tilde{\partial}_k\Bigg( \hspace{-6.0cm}\includegraphics[bb = -140 528 250 0,scale=0.5,angle=0]{vertex2.pdf}\hspace{1cm}\Bigg)&=&g^3Z_{g,k}^3Z_{A,k}^{1/2}Z_{c,k}\int_q k\tilde{\partial}_k \frac{1}{q_R^2}\frac{1}{((q-p)_R^2)^2}\nonumber\\
&& \quad \times \Big(\delta_{mn}-(1-\xi_k)\frac{(q-p)_n(q-p)_m}{(q-p)^2}\Big) \nonumber\\
&& \quad \times \Big(\delta_{or}-(1-\xi_k)\frac{(q-p)_o(q-p)_r}{(q-p)^2}\Big)f^{fdc}f^{aef}f^{bde}q_n p_r \nonumber\\
&& \quad \times \Big(\delta_{im}(q-p)_o+\delta_{mo}(2p-2q)_i+\delta_{io}(q-p)_m\Big)\nonumber\\
&=&p_if^{abc}g^3Z_{g,k}^3Z_{A,k}^{1/2}Z_{c,k}C_2(A)\Omega_d k^{d-4}\frac{3(1-d)\xi_k}{d(d+2)}+{\cal O}(p^2)\,,
\eea
where 
\bea
\label{Eq:qR}
q_R^2 \equiv q^2+R_k(q) = q^2 + (k^2-q^2)\Theta(k^2-q^2)\,,
\eea
and $\Omega_d = [2^{d-1} \pi^{d/2}\Gamma(d/2)]^{-1}$ is the area of a $d$-dimensional unit sphere divided by the factor of $(2\pi)^d$. 
In the calculations above, in the right-hand sides of (\ref{Eq:vertexd1}) and (\ref{Eq:vertexd2}) we neglected the $k$-dependence of the rescaling factors $Z_{g,k}$, $Z_{A,k}$, and $Z_{c,k}$, and used that $f^{abc}f^{abd}=C_2(A)\delta^{cd}$, where $C_2(A)$ is the value of the Casimir operator $\hat{T}_i\hat{T}_i$ in the adjoint representation. We also made use of the identity $f^{oaf}f^{fbe}f^{eco} = -\frac{1}{2} C_2(A) f^{abc}$. Collecting (\ref{Eq:vertexd1}) and (\ref{Eq:vertexd2}), (\ref{Eq:vertex}) leads to
\bea
\label{Eq:vertexfinal}
\frac{1}{Z_{g,k}Z_{A,k}^{1/2}Z_{c,k}}k\partial_k (Z_{g,k}Z_{A,k}^{1/2}Z_{c,k})=g^2Z_{g,k}^2C_2(A)\Omega_d k^{d-4}\frac{-3\xi_k}{d+2}\,.
\eea

\subsection{Ghost wavefunction renormalization}
Concerning the flow of $Z_{c,k}$, we only need to calculate one diagram, for which we have
\bea
p^2\delta^{ab}k \partial_k Z_{c,k}&=&k\tilde{\partial}_k\Bigg( \hspace{-6.0cm}\includegraphics[bb = -50 522 250 0,scale=0.7,angle=0]{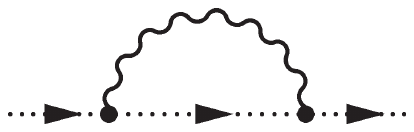}\hspace{1.7cm}\Bigg)\nonumber\\
&=&g^2Z_{g,k}^2Z_{c,k}f^{adc}f^{bdc}\int_q k\tilde{\partial}_k\frac{1}{q_R^2}\frac{-1}{(q-p)_R^2}\Big(\delta_{ij}-(1-\xi_k)\frac{(q-p)_i(q-p)_j}{(q-p)^2}\Big)q_ip_j \nonumber\\
&=&g^2Z_{g,k}^2Z_{c,k}C_2(A)p^2\delta^{ab}\int_{|q|<k} \frac{-2}{k^4}\big(-1+(\xi_k-1)(1-2x^2)\big)+{\cal O}(p^3)\,,
\eea
where $x=\hat{p} \cdot \hat{q}$ is the dot product of unit vectors along momenta $p$ and $q$, and again we neglected the $k$-dependence of the rescaling factors in the right-hand side. Performing the integral, we arrive at
\bea
\label{Eq:Zckfinal}
\frac{k\partial_k Z_{c,k}}{Z_{c,k}}=g^2Z_{g,k}^2C_2(A)\Omega_d k^{d-4}\frac{4(d-1)-2\xi_k(d-2)}{d^2}\,.
\eea

\subsection{Gauge wavefunction renormalization}
The flow of $Z_{A,k}$ turns out to be the most complicated of all. In principle, we have the following five diagrams, which need to be expanded, again, up to ${\cal O}(p^2)$: 
\bea
\label{Eq:ZAalldiag}
\hspace{-1.0cm}k\partial_k \Big[Z_{A,k}\big(p^2\delta_{ij}-p_ip_j(1-\xi_k^{-1})\big)\Big]\delta^{ab}=k\tilde{\partial}_k&&\hspace{-0.1cm}\Bigg( \hspace{-6.0cm}\includegraphics[bb = -160 508 250 0,scale=0.5,angle=0]{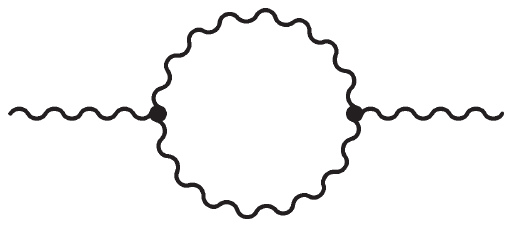}\hspace{1.45cm}+\hspace{-6.0cm}\includegraphics[bb = -160 508 250 0,scale=0.5,angle=0]{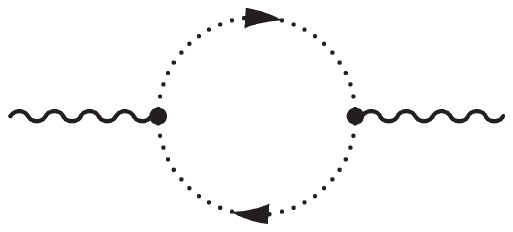}\hspace{1.5cm}+\nonumber\\
&&\hspace{-8.0cm}\includegraphics[bb = -160 508 250 0,scale=0.53,angle=0]{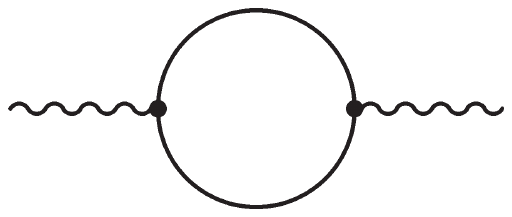}\hspace{1.7cm}+\hspace{-6.5cm}\includegraphics[bb = -150 535 250 0,scale=0.52,angle=0]{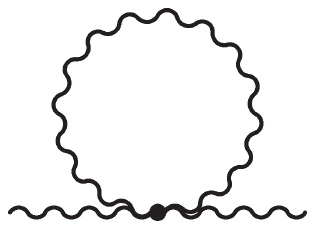} \hspace{1.1cm} +\hspace{-6.45cm}\includegraphics[bb = -150 535 250 0,scale=0.52,angle=0]{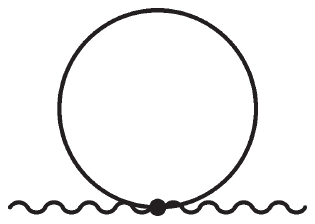} \hspace{1.1cm}\Bigg)\,.
\eea
Note that, the last two of them are $p$ independent, and they should cancel against similar contributions coming from the previous diagrams. This, however, does not happen in the FRG framework. The reason is that the regulator function explicitly breaks the gauge symmetry and the vanishing gluon mass, $m_{A,k}=0$, is not protected anymore. This makes no problem, as once $m_{A,\Lambda}$ is  appropriately adjusted at the ultraviolet (UV) scale, it vanishes as $k \rightarrow 0$, and so it can be considered as an irrelevant operator. Therefore, only considering the first three diagrams, up to $p$ independent constants, they read
\bea
\label{Eq:ZAdiag1}
k\tilde{\partial}_k&&\hspace{-6.5cm}\includegraphics[bb = -160 508 250 0,scale=0.5,angle=0]{ZA1.pdf}\hspace{1.5cm}=g^2Z_{g,k}^2 Z_{A,k}f^{adc}f^{bdc}\frac12 \int_q k\tilde{\partial}_k \frac{1}{q_R^2}\frac{1}{(q+p)_R^2}\nonumber\\
&&\hspace{1.4cm}\times\Big(\delta_{mn}-(1-\xi_k)\frac{q_mq_n}{q^2}\Big)\Big(\delta_{ol}-(1-\xi_k)\frac{(q+p)_o(q+p)_l}{(q+p)^2}\Big)\nonumber\\
&&\hspace{1.4cm}\times \Big(\delta_{im}(p-q)_o+\delta_{mo}(2q+p)_i+\delta_{io}(-2p-q)_m\Big)\nonumber\\
&&\hspace{1.4cm}\times \Big(\delta_{jl}(2p+q)_n+\delta_{ln}(-2q-p)_j+\delta_{nj}(q-p)_l\Big)\nonumber\\
&&\hspace{1.4cm}=g^2Z_{g,k}^2 Z_{A,k}\delta^{ab}C_2(A)\int_{|q|<k}\frac{-2}{k^4(q+p)_R^2}\bigg(4(1-d)q_iq_j-2\xi_k(\delta_{ij}q^2-q_iq_j) \nonumber\\
&&\hspace{1.4cm}+p^2\Big[8(x^2-1)\delta_{ij}+(9-d)\hat{p}_i\hat{p}_j+4\hat{q}_i\hat{q}_j -4x(\hat{p}_i\hat{q}_j+\hat{p}_j\hat{q}_i)-4x^2\hat{q}_i\hat{q}_j\nonumber\\
&&\hspace{1.4cm}+\xi_k\Big(3\delta_{ij}-3\hat{p}_i\hat{p}_j-4\hat{q}_i\hat{q}_j
+4x(\hat{p}_i\hat{q}_j+4\hat{p}_j\hat{q}_i)-8x^2\delta_{ij}+4\hat{q}_i\hat{q}_jx^2\Big)\Big]\bigg)+{\cal O}(p^3) \nonumber\\
&=&g^2Z_{g,k}^2Z_{A,k}\delta^{ab}C_2(A)\Omega_d k^{d-4} \Bigg[\frac{12d(d+1)-40}{d^2(d+2)}\Big(p^2\delta_{ij}-\frac{d(6+11d-d^2)-24}{6d(d+1)-20}p_ip_j\Big)\nonumber\\
&&\hspace{3.9cm}+\xi_k\frac{40+2d(5-4d)}{d^2(d+2)}\Big(p^2\delta_{ij}-\frac{3(d^2-8)}{d(4d-5)-20}p_ip_j\Big)\Bigg]+{\cal O}(p^3)\,,\nonumber\\
\eea
\bea
\label{Eq:ZAdiag2}
\hspace{-0.5cm}k\tilde{\partial}_k \hspace{-5.9cm}\includegraphics[bb = -160 508 250 0,scale=0.5,angle=0]{ZA2.pdf}\hspace{1.5cm}&=&g^2Z_{g,k}^2Z_{A,k}f^{adc}f^{bdc}\int_q k\tilde{\partial}_k\frac{1}{q_R^2}\frac{1}{(q+p)_R^2}(p+q)_iq_j\nonumber\\
&=&g^2Z_{g,k}^2Z_{A,k}\delta^{ab}C_2(A)\Omega_d k^{d-4}\frac{2}{d(d+2)}(p^2\delta_{ij}+2p_ip_j)+{\cal O}(p^3),\nonumber\\
\eea
\bea
\label{Eq:ZAdiag3}
\hspace{-0.5cm}k\tilde{\partial}_k\hspace{-5.9cm}\includegraphics[bb = -160 508 250 0,scale=0.5,angle=0]{ZA3.pdf}\hspace{1.5cm}&=&-g^2Z_{g,k}^2Z_{A,k}\Tr[\hat{T}^a\hat{T}^b] N_{\rm f} \int_q k\tilde{\partial}_k \frac{1}{q_R^2}\frac{1}{(q+p)_R^2}(p+2q)_i(p+2q)_j\nonumber\\
&=&g^2Z_{g,k}^2Z_{A,k}\delta^{ab}\Omega_d k^{d-4}\frac{-4N_{\rm f}}{d(d+2)}\Big(p^2\delta_{ij}-\frac{d-2}{2}p_ip_j\Big)+{\cal O}(p^3)\,.
\nonumber \\
\eea
Similarly to the previously employed approximation, we once again did not let $\tilde{\partial}_k$ act on the rescaling factors, and note that we implicitly assumed that the scalar mass is (approximately) zero, as we are working at the critical temperature. Collecting (\ref{Eq:ZAdiag1}), (\ref{Eq:ZAdiag2}) and (\ref{Eq:ZAdiag3}), (\ref{Eq:ZAalldiag}) takes the following form:
\bea
\label{Eq:ZAalldiagsum}
\hspace{0cm}k\partial_k \Big[Z_{A,k}\big(p^2\delta_{ij}-p_ip_j(1-\xi_k^{-1})\big)\Big]&=&g^2Z_{g,k}^2Z_{A,k}\Omega_d k^{d-4}\nonumber\\
&&\hspace{-3.4cm}\times\Big[p^2\delta_{ij}\Big(-\frac{4N_{\rm f}}{d(d+2)}+C_2(A)\frac{12d^2+14d-40-2\xi_k(4d^2-5d-20)}{d^2(d+2)}\Big) \nonumber\\
&&\hspace{-3.2cm}-p_ip_j\Big(-\frac{2N_{\rm f}(d-2)}{d(d+2)}+C_2(A)\frac{-2d^3+22d^2+8d-48-6\xi_k(d^2-8)}{d^2(d+2)}\Big)\Big]\,. \nonumber\\
\eea
For $d=4$, what we see is that the right-hand side of (\ref{Eq:ZAalldiagsum}) is transverse, which means that the gauge-fixing parameter has to obey $\xi_k=\const \times Z_{A,k}$, as expected from perturbation theory. However, if $d\neq 4$, then the longitudinal projection of the gluon propagator also flows. Luckily, this can be compensated by choosing $\xi_k$ appropriately, but this also means that the gauge-fixing parameter has to be tailored to the approximation in question and is not arbitrary. First we read off from the transverse component that
\bea
\label{Eq:ZAkfinal}
\frac{k\partial_k Z_{A,k}}{Z_{A,k}}=g^2Z_{g,k}^2\Omega_d k^{d-4}\Big(-\frac{4N_{\rm f}}{d(d+2)}+C_2(A)\frac{12d^2+14d-40-2\xi_k(4d^2-5d-20)}{d^2(d+2)}\Big)\,,\nonumber\\
\eea
and then by assuming that $\xi_k\equiv \xi$ is scale independent, the longitudinal projection leads to the following consistency condition:
\bea
k\partial_k Z_{A,k}(1-\xi^{-1})&=&g^2Z_{g,k}^2Z_{A,k}\Omega_d k^{d-4}\nonumber\\
&&\times\Big(-\frac{2N_{\rm f}(d-2)}{d(d+2)}+C_2(A)\frac{-2d^3+22d^2+8d-48-6\xi(d^2-8)}{d^2(d+2)}\Big)\,,\nonumber\\
\eea
which is equivalent to
\bea
\label{Eq:consistency}
\Bigg[-\frac{4N_{\rm f}}{d+2}+C_2(A)\frac{12d^2+14d-40-2\xi(4d^2-5d-20)}{d(d+2)}\Bigg]\Big(1-\xi^{-1}\Big) \nonumber\\
= -\frac{2N_{\rm f}(d-2)}{d+2}+C_2(A)\frac{-2d^3+22d^2+8d-48-6\xi(d^2-8)}{d(d+2)}\,.
\eea

\subsection{$\beta$ function of the gauge coupling}
Now we are in a position to obtain the $\beta$ function of the gauge coupling. Plugging (\ref{Eq:vertexfinal}), (\ref{Eq:Zckfinal}) and (\ref{Eq:ZAkfinal}) to (\ref{Eq:betagdef}), we get
\bea
\label{Eq:betafinal}
\beta(g)=-(4-d)\bar{g}_k-\bar{g}_k^3\Omega_d\Bigg[C_2(A)\frac{10d^2+11d-28+
\xi_k(-3d^2+5d+28)}{d^2(d+2)}-\frac{2N_{\rm f}}{d(d+2)}\Bigg]\,. \nonumber\\
\eea
In $d=4$, we get back the usual one-loop perturbative result,
\bea
\label{Eq:betagdim4}
\beta(g)|_{d=4}=-\frac{\bar{g}_k^3}{(4\pi)^2}\Big(\frac{11}{3}N_{\rm c}-\frac{N_{\rm f}}{6}\Big)\,,
\eea
where we used that $C_2(A)=N_{\rm c}$ for $\SU(N_{\rm c})$. It can be argued that only specifically in $d=4$, this result is regulator independent, and one could have used any other profile function for $R_k$.
Note that if one is to look for reproducing higher-order contributions in terms of $g$ in perturbation theory or improve upon the present FRG scheme, one needs to take into account the $k$-dependence of the rescaling factors in the right-hand sides of the individual flow equations, which, as stressed earlier, has been neglected here.

For $d=3$, (\ref{Eq:betafinal}) gives the following new result:
\bea
\label{Eq:betagdim3}
\beta(g)|_{d=3}=-\bar{g}_k-\frac{\bar{g}_k^3}{2\pi^2}\Bigg[\Bigg(\frac{19}{9}+\frac{16}{45}\xi\Bigg)N_{\rm c}-\frac{2N_{\rm f}}{15} \Bigg]\,.
\eea
One notes that, similarly to the Abelian case \cite{fejos16,fejos17}, $\xi$ remains in the expression of $\beta(g)$. Here $\xi$ must be fixed such that it satisfies the consistency condition (\ref{Eq:consistency}). The latter has two solutions for $\xi$:
\bea
\label{Eq:xichoice}
\xi_\pm = 1 + \frac{N_{\rm f}}{4} \pm \frac14\sqrt{N_{\rm f}^2 - 8 N_{\rm f} + 456}\,.
\eea
Note that $\xi_- \rightarrow 2 + {\cal O}(1/N_{\rm f})$, while $\xi_+ \rightarrow N_{\rm f}/2 + {\cal O}(1/N_{\rm f})$, as $N_{\rm f}\rightarrow \infty$. We require that at least for $N_{\rm f}\rightarrow \infty$, where the contributions of non-Abelian gauge fields is suppressed compared with those of matter fields, the $\beta$ function admits a nontrivial fixed point similarly to the Abelian Higgs model. This rules out the possibility of using $\xi_+$, but allows for using $\xi_-$. Then, it turns out  that, if $N_{\rm c}=3$, for $N_{\rm f} \leq 55$ we find that $\beta(g)|_{d=3}<0$ for all $\bar{g}>0$, and hence, it does not admit any IR fixed point in $d=3$. This rules out the possibility of having a second-order color superconducting phase transition.

Note that, we did not have to calculate any RG flow in the scalar sector, i.e. the RG flows of $\alpha$, $\beta_1$, $\beta_2$ in (\ref{GL}). Even though if there were scaling solutions of them for $\bar{g}=0$, they could never be IR stable as $\beta(g) < 0$. This shows that if $N_{\rm f} \leq 55$, the Ginzburg-Landau potential $V(\phi)$ has no relevance in the order of the transition, and it is always of first order. This is in contrast with ordinary superconductivity, where the gauge coupling does admit a nontrivial solution, which when plugged into the flows of the scalar sector, does produce an IR stable fixed point \cite{fejos16,fejos17}. Gluon fluctuations, however, are stronger than those of the photon, and in non-Abelian gauge theories the latter scenario is not realized.

\section{Vertex regulators}
\label{sec:vertex}
Although the calculations in the previous section represent a perfectly legitimate regularization of the continuum theory, one point needs mentioning. As argued already in \cite{fejos19}, there is another choice of regularization with the same profile function $R_k(q)$, once one is encountered with momentum-dependent vertices. A general problem of momentum-dependent vertices is that, strictly speaking, they are inconsistent in the sense that they do not follow the value of the regulated momenta of the lines attached to them, once the loop momentum becomes $q<k$. Even though in $d=4$, all the $\beta$ functions must be universal (at least at the lowest order), and thus this issue will never have any effect, it certainly changes results in other dimensions. In the Abelian-Higgs model, it has been shown that regulating momentum-dependent vertices is of particular importance regarding the cancellation of gauge-dependent contributions in self-coupling flows \cite{fejos19}. We, therefore, in this section perform another line of calculation by (partially) taking into account vertex regularization.

A severe problem of vertex regularization is that the FRG regulator, by definition, is quadratic in the fields, and the vertices are at least cubic. Therefore, one is not allowed to introduce any regulator at the level of the vertices, as it violates the structure of $\Gamma_k$. There are special cases, however, when quadratic regulators can indeed regulate vertices. Imagine a diagram, where an external line with zero momentum can be attached to a momentum-dependent three-point vertex. This construction always emerges in the functional integral from a term, where the field corresponding to the external line is a constant, referring to a background field value, where the effective action is evaluated. Then, by adding background-dependent off-diagonal components to the regulator matrix, one can indeed switch the loop momenta in the vertex $q_i \rightarrow q_{R,i}$, where $q_{R,i}=q_i+(\hat{q}_ik-q_i)\Theta(k^2-q^2)$. We stress once again that, this only works when the vertex depends only on a single momentum variable, and the background-dependent off-diagonal regulator is introduced for those two fields, along which this momentum is flowing through. Although none of the diagrams we have been involved with in the previous section contains vertices with only a single momentum flow, we perform an expansion in the external momentum in all the diagrams, which formally may leave the remaining structure suitable for the vertex regularization described above. 

Concerning the gauge-antighost-ghost vertex and the ghost wavefunction renormalization, the procedure works. It is indeed true that after expanding the corresponding diagrams around the external momentum, the coefficient can be regarded as an expression, where vertices satisfy the aforementioned property of the single momentum flow. Concerning the scale dependence of the gauge wavefunction renormalization, however, the procedure fails. The reason is that in the flow of $Z_{A,k}$ there is always a contribution at ${\cal O}(p^2)$ that is coming from the $p$ dependence of the regulated propagator, i.e. there are always some vertices that involve two momenta (external and loop), and thus the trick above does not apply. It would be interesting to come up with a method that solves the problem of the vertex regularization in these type of diagrams, but we leave it for further studies.

In what follows, we analyze how the gauge-antighost-ghost vertex and the ghost wavefunction renormalization change in the new regularization procedure. For the former one, we need to analyze once again (\ref{Eq:vertexd1}) and (\ref{Eq:vertexd2}). When we switch $q \rightarrow q_R$ in the vertices, one effectively changes the momentum integral $\int_q k\tilde{\partial}_k [q^2/(q_R^2)^3] \rightarrow \int_q k\tilde{\partial}_k [1/(q_R^2)^2]$. This results in a factor of $\frac{2(d+2)}{3d}$ compared to (\ref{Eq:vertexd1}) and (\ref{Eq:vertexd2}). Therefore, (\ref{Eq:vertexfinal}) changes to
\bea
\label{Eq:vertexfinalnew}
\frac{1}{Z_{g,k}Z_{A,k}^{1/2}Z_{c,k}}k\partial_k (Z_{g,k}Z_{A,k}^{1/2}Z_{c,k})=g^2Z_{g,k}^2C_2(A)\Omega_d k^{d-4}\frac{-2\xi_k}{d}\,.
\eea
As expected, for $d=4$, there is no difference, but for any other dimension, the RG flow changes. A similar analysis for the ghost wavefunction renormalization changes (\ref{Eq:Zckfinal}) to
\bea
\label{Eq:Zckfinalnew}
\frac{k\partial_k Z_{c,k}}{Z_{c,k}}=g^2Z_{g,k}^2C_2(A)\Omega_d k^{d-4}\frac{3-\xi_k}{d}\,,
\eea
which again shows the expected universality in $d=4$. The flow of $Z_{A,k}$ is unchanged because of the reasons explained above, thus, using (\ref{Eq:betagdef}), (\ref{Eq:ZAkfinal}), (\ref{Eq:vertexfinalnew}) and (\ref{Eq:Zckfinalnew}), the modified $\beta(g)$ function becomes
\bea
\beta(g)=-(4-d)\bar{g}_k-\bar{g}_k^3\Omega_d\Bigg[N_{\rm c}\frac{9d^2-13d+20+\xi_k(3d^2-7d-20)}{d^2(d+2)}-\frac{2N_{\rm f}}{d(d+2)}\Bigg]\,,
\eea
where we again used $C_2(A)=N_{\rm c}$ for $\SU(N_{\rm c})$.
In $d=4$ it reduces to the usual form (\ref{Eq:betagdim4}), and in $d=3$ we get
\bea
\label{Eq:betagdim3new}
\beta(g)=-\bar{g}_k-\frac{\bar{g}_k^3}{2\pi^2}\Bigg[\Bigg(\frac{20}{9}+\frac{14}{45}\xi_k\Bigg)N_{\rm c}-\frac{2N_{\rm f}}{15}\Bigg]\,.
\eea
This approximation is clearly different from that of (\ref{Eq:betagdim3}), but the numerical factors are very close to each other. When we solve (\ref{Eq:consistency}) for $\xi$ and choose, again, $\xi_-$ [see (\ref{Eq:xichoice})] we once again find that $\beta(g)<0$, and thus no IR fixed point can appear, where $\bar{g}\neq 0$, and the phase transition is (presumably) of first order.

\section{Conclusions}
\label{sec:conclusion}
In this paper we have analyzed if the Ginzburg-Landau effective theory of color superconductivity can admit IR stable fixed points, which could describe a possible second-order transition in the system. We have employed the functional renormalization group (FRG) approach, since it provides a framework, where the $\beta$ functions of the couplings can be evaluated directly in $d=3$. This is of particular importance, since the $\epsilon$ expansion around $d=4$ is expected to break down in gauge theories with scalar fields; see, e.g. the case of ordinary superconductivity \cite{kleinert06}.

We have calculated the $\beta$ function of the gauge coupling, and found that irrespectively of the structure of the scalar potential, for $N_{\rm c}=3$ and $N_{\rm f} \leq 55$, the flow equation can never admit an IR stable fixed point solution in $d=3$. This is an indirect evidence that the color superconducting phase transition can only be of first order. The existence of this first-order phase transition can be traced back to the celebrated property of non-Abelian gauge theories --- the asymptotic freedom --- according to which the only existing fixed point is in the UV and is trivial. 

Our study is relevant not only from the point of view of color superconductivity, but it helps facilitate a deeper understanding of how the FRG works for non-Abelian gauge theories. We have shown that, as expected from the earlier Abelian study \cite{fejos17}, gauge symmetry violation by the IR regulator seems to be causing incompatibility between the flow equation and the structure of the (scale-dependent) effective action. The main problem was that as fluctuations are integrated out, longitudinal projection of the gluon propagator starts to flow with the RG scale. We have shown that, this can be compensated by choosing appropriate value for the gauge-fixing parameter $\xi$, which is in line with the expectations coming from the Abelian-Higgs model.

We have also touched upon the problem of vertex regularization, which is necessary to be dealt with from the point of view of RG consistency, once one encounters with momentum-dependent vertices. We have found a way to regularize these vertices when calculating the scale dependence of the gauge-antighost-ghost vertex and the ghost wavefunction renormalization $Z_{c,k}$, but our method is inapplicable for the gluon wavefunction renormalization $Z_{A,k}$. We have shown that the $\beta$ function of the gauge coupling does change with respect to the aforementioned extension of the regularization procedure, but it is only minor and does not affect the main result. For a more complete picture, it would be interesting to come up with a solution of regularizing momentum-dependent vertices in the calculation of the flow of $Z_{A,k}$. 

Our method here could also be extended to non-Abelian gauge theories with fermions. In order to understand the realistic QCD phase diagram at intermediate density region, it would be important to extend the present FRG approach to Ginzburg-Landau theories incorporating the effects of finite quark masses \cite{iida03,iida04} and the competition between chiral symmetry breaking and color superconductivity \cite{hatsuda06,yamamoto07}. Finally, the method could also have relevance in regard to the electroweak phase transition \cite{kajantie96}. These issues represent future works to be reported elsewhere.

\section*{Acknowledgements}
This work was supported by the Keio Institute of Pure and Applied Sciences (KiPAS) project in Keio University and MEXT-Supported Program for the Strategic Research Foundation at Private Universities, ``Topological Science'' (Grant No.~S1511006). G.~F. was also supported by the Hungarian National Research, Development and Innovation Office (Project No. 127982), the J\'anos Bolyai Research Scholarship of the Hungarian Academy of Sciences, and the \'UNKP-19-4 New National Excellence Program of the Ministry for Innovation and Technology of Hungary. N.~Y. was supported by JSPS KAKENHI Grant No.~19K03852.

\renewcommand{\theequation}{A\arabic{equation}}

\appendix
\section{Common integrals}
\label{sec:integral}
In this appendix we show some examples how to extract the ${\cal O}(p^2)$ term from several integrals that appear in the main text. 
As an example, we first consider the following integral:
\bea
I_n = -\int_q k\tilde{\partial}_k \frac{q^{2n}}{q_R^2(p+q)_R^2}\,,
\eea
where $n=1,2,\dots$ and $q_R^2$ is defined in (\ref{Eq:qR}). Here, the differentiation with respect to $k$ acts on $1/q_R^2$ and $1/(p+q)_R^2$,
and by performing the change of variables $q \rightarrow -q - p$ in the latter contribution, their summation can be written as
\bea
I_n 
&=& \int_q \bigg(-k\tilde{\partial}_k \frac{1}{q_R^2}\bigg)\frac{1}{(p+q)_R^2}\Big(q^{2n}+(p+q)^{2n}\Big) \nonumber \\
&=&\int_{q_+<|q|<k} \frac{2}{k^2}\Big(\frac{1}{(p+q)^2}-\frac{1}{k^2}\Big)\Big(q^{2n}+(p+q)^{2n}\Big)\Theta(x) \nonumber\\
& &+\int_{|q|<k}\frac{2}{k^4}\Big(q^{2n}+(p+q)^{2n}\Big)+{\cal O}(p^3)\nonumber\\
&=&\frac{4}{d+2n}\Omega_dk^{d+2n-4}+\frac{2(n-1)(2dn+d-2)}{d(d+2n-2)} \Omega_d k^{d+2n-6} p^2 +{\cal O}(p^3)\,.
\eea
where $q_+=k-px+p^2(x^2-1)/(2k)+{\cal O}(p^3)$. In the second line above, we used that $(p+q)_R^2=k^2$ if $q<q_+$ and $(p+q)_R^2=k^2$ if $q_+<q<k$, and we assumed that $\hat{p} \cdot \hat{q}\equiv x >0$. (If $x<0$, then obviously $(p+q)_R^2\equiv k^2$ for all $q<k$.)  In the last line, we used that under the angular integral one may substitute $x^2\rightarrow 1/d$, $\Theta(x)\rightarrow 1/2$.

Similarly, we can derive the following useful integrals:
\bea
-\int_q k\tilde{\partial}_k \frac{p_iq_j}{q_R^2(q+p)_R^2} = -\int_{|q|<k} \frac{2p_i p_j}{k^2(q+p)_R^2}  =- \frac{2}{d} \Omega_d k^{d-4}p_ip_j + {\cal O}(p^3)\,,
\eea
\bea
\hspace{-0.5cm}-\int_q k\tilde{\partial}_k \frac{q_iq_j}{q_R^2(p+q)_R^2}&=&-\int_q \bigg(k\tilde{\partial}_k\frac{1}{q_R^2}\bigg)\frac{1}{(q+p)_R^2}\Big(q_iq_j+(p+q)_i(p+q)_j\Big)\nonumber\\
&=&\int_{|q|<k}\frac{2}{k^4}(2q_iq_j+p_ip_j)\nonumber\\
& &+\int_{q_+<|q|<k} \frac{2\Theta(x)}{k^2}\bigg[\frac{1}{(q+p)^2}-\frac{1}{k^2}\bigg]\Big(q_iq_j+(q+p)_i(q+p)_j\Big) \nonumber\\
&=&\frac{4\delta_{ij}}{d(d+2)}\Omega_d k^{d-2} - \frac{2}{d(d+2)} \Omega_d  k^{d-4} \Big(p^2\delta_{ij}-dp_i p_j\Big)+{\cal O}(p^3)\,.
\eea
During the calculations we employed the angular integral identities,
\bea
\int_\Omega \ \hat{q}_i\hat{q}_j = \frac{1}{d}\Omega_d\delta_{ij}\,, \quad 
\int_\Omega \ \hat{q}_i\hat{q}_j\hat{q}_k\hat{q}_l = \frac{1}{d(d+2)}\Omega_d (\delta_{ij}\delta_{kl}+\delta_{ik}\delta_{jl}+\delta_{il}\delta_{jk})\,.
\eea

\end{document}